\documentstyle[sprocl]{article}

\def\be{\begin{equation}}
\def\ee{\end{equation}}
\def\bea{\begin{eqnarray}}
\def\eea{\end{eqnarray}}
\begin{document}

\title{Is SU(3) Chiral Perturbation Theory an Effective Field Theory?}

\author{Barry R. Holstein}

\address{Department of Physics and Astronomy, University of
Massachusetts, Amherst,\\ MA  01003, USA\\E-mail: holstein@phast.umass.edu}
%%%%%%%%%%%%%%%%%%%%%%%%%%%%%%%%%%%%%%%%%%%%%%%%%%%%%%%%%%%%%%
% You may repeat \author \address as often as necessary      %
%%%%%%%%%%%%%%%%%%%%%%%%%%%%%%%%%%%%%%%%%%%%%%%%%%%%%%%%%%%%%%

\maketitle\abstracts{We argue that the difficulties associated with the
convergence
properties of conventional SU(3) chiral perturbation theory can be ameliorated
by
use of a cutoff, which suppresses the model-dependent short distance effects in
such calculations.}

\noindent
\section{Introduction}
Tony Thomas presented us with a challenge in setting up this workshop---to make
the
connection between quark and hadronic degrees of freedom.
I have accepted this challenge and will go even further by attempting to make
contact between the {\it physics} of Tony Thomas and that of Ulf Meissner!
My point of contact will be the study of low energy baryon phyics, studied a
number
of years ago by Tony and his collaborators in the context of the cloudy bag
model
and examined in recent years by Ulf and his collaborators using the technique
of
chiral perturbation theory.

Let's begin with some simple baryon phenomenology which was studied in the
1960's
via the quark model, but which has now been updated via SU(3) chiral
perturbative
studies.  In each case the simple quark model picture is reasonably successful
but
the addition of chiral loops, which are supposed to be essentially model
independent,
destroys experimental agreement:

\begin{itemize}
\item [i)] Baryon masses, wherein the Gell-Mann-Okubo relation obtains
at first order in quark mass and is well satisfield experimentally.
Chiral loops, however, make significant---{\cal
O}(50-100\%)---corrections to individual masses.\cite{1}

\item [ii)] Semileptonic hyperon decay, wherein a simple SU(3)
representation of the axial couplings in terms of F,D couplings yields
an excellent fit to experiment.  Chiral loops make {\cal O}(30-50\%)
corrections to individual couplings and destroy this agreement.\cite{2}

\item [iii)] Nonleptonic hyperon decay, wherein a basic SU(3) fit to
s-wave amplitudes provides an excellent representation of the
experimental numbers in terms of f,d couplings.  Chiral loops make
{\cal O}(30-50\%) corrections to individual terms and destroy this
agreement.  (The situation is somewhat more confused in the case of the
p-waves wherein a significant cancellation between pole terms exists
at lowest order and the validity of the chiral expansion is suspect.\cite{3})

\item [iv)] Magnetic moments, wherein an SU(3) representation (especially with
the strange quark mass accounted for) provides a reasonable fit to measured
quantities, but where loop corrections make {\cal O}(50-90\%) corrections,
which
invalidate the goodness of fit.\cite{4}
\end{itemize}
Of course, the disagreement brought about by such loop modifications can be
eliminated by contributions from yet higher order terms in the chiral
expansion.
However, this gives one concern about the convergence properties of the series.
Thus I would answer the question raised in my title---yes it is an effective
field
theory but it is not an {\it effective} field theory!

Below we shall make the case that this problem can be solved by using a cutoff
regularization in order to calculate chiral corrections rather than the
conventional
dimensional regularization.  The important point here is that a significant
component
of the loop correction arises from meson propagation over distances small
compared
to the $\sim$ 1 fm size of a typical baryon.  This piece is not only large but
is
model dependent and unphysical---the representation of the physics in terms of
point particles misrepresents the short distance physics.  We shall develop a
procedure whereby only the---model independent---long distance component is
retained
and will show how this resolves the problem with respect to the convergence of
the
chiral expansion.

In the next section we shall make a somewhat detailed analysis of the problem
of
baryon mass renormalization, and then in section III we will extend our
discussion
to include the other aspects of low energy baryon physics referred to above.  A
short summary appears in a concluding section IV.

\section{Baryon Mass}

An example of the difficulties of baryon chiral perturbation
theory is provided by the analysis of baryon masses. Such masses
have an expansion in the masses of the quarks ($m_{q}$),
or equivalently in terms of the pseudoscalar meson masses ($m_{M}$)
\begin{eqnarray}
M_B & = & M_0 + \Sigma_q ~\bar{b}_q m_{q} + \Sigma_q ~\bar{c}_q m_{q}^{3/2}
 + \Sigma_q ~\bar{d}_q m_{q}^2 + ...  \\
 & = & M_0 + \Sigma_M ~{b}_M m_{M}^2 + \Sigma_M ~{c}_M m_{M}^{3}
 + \Sigma_M ~{d}_M m_{M}^4 + ...
\end{eqnarray}
\noindent Here $M_0$ is a common mass and $b_M$ and $d_M$ contain adustable
parameters representing terms in the effective Lagrangian. However,
the non-analytic $m_q^{3/2}$ terms come from loop diagrams, and
the coefficients are not adjustable---they are given in terms of
the baryon-meson coupling constants. The leading SU(3) breaking
terms involving $b_M$ go back to Gell-Mann and Okubo\cite{5}, the non-analytic
corrections from one-loop diagrams, represented above by $c_M$,
were first calculated by Langacker and Pagels\cite{6}, and the complete
$m_M^4$ corrections (including diagrams up to two loops) were
calculated by Borasoy and Meissner\cite{1}. The convergence difficulties
in the expansion are demonstrated by the resulting fit for the
nucleon mass where, in the same sequence, the different
contributions are given, in GeV, by\cite{1}
\begin{equation}
M_N = 0.711 + 0.202 - 0.272 + 0.298+\ldots
\end{equation}
or, more dramatically for the $\Xi$,
\begin{equation}
M_\Xi = 0.767 +0.844  - 0.890 + 0.600+\ldots
\end{equation}
\noindent The non-analytic terms appear unavoidably large and the
expansion has certainly not converged at this order. The
final fit also violates the Gell-Mann-Okubo relation by an amount that
is five times larger than the experimentally observed value.

To one-loop order, the explicit form of the contributions to
the baryon masses is given by\cite{7}
\begin{eqnarray}
M_N & = & {\hat M}_0 - 4m_K^2 b_D + 4(m_K^2 - m_\pi^2) b_F + L_N  \nonumber\\
M_\Lambda & = &  {\hat M}_0-{4 \over 3}(4 m_K^2 - m_\pi^2) b_D +
L_\Lambda \nonumber\\
M_\Sigma & = &  {\hat M}_0 - 4 m_\pi^2 b_D + L_\Sigma \nonumber\\
M_\Xi & = &  {\hat M}_0 - 4m_K^2 b_D - 4(m_K^2 - m_\pi^2) b_F + L_\Xi
\end{eqnarray}
where
\begin{equation}
 {\hat M}_0 = M_0 - 2(2 m_K^2 + m_\pi^2) b_0
\end{equation}
with $M_0$, $b_D$, $b_F$ and $b_0$ as free parameters. (Note that
$M_0$ and $b_0$ do not have separate effects, but only enter in the
combination ${\hat M}_0$.) The ingredients $L_B$ contain the
nonanalytic contributions from loop diagrams, and have the form
\begin{equation}
L_i=-{1\over 24\pi F_\pi^2}\sum_j\kappa^j_im_j^3
\end{equation}
where $\kappa_i^j$ are given in terms of the $D$ and $F$ baryon axial-vector
couplings
in ref. 1.
These non-analytic terms are quite large, having values
\begin{eqnarray}
L_N & = & -0.31 ~GeV, \nonumber\\
L_\Lambda & = & -0.66 ~GeV, \nonumber\\
L_\Sigma & = & -0.67 ~GeV, \nonumber\\
L_\Xi & = & -1.02 ~GeV.
\end{eqnarray}
using $D = 0.806$, $F = 0.46$ and $F_\pi = 93$MeV.
In particular, the $\Xi$ mass shift is
clearly unphysically large.
It is not possible to obtain a reasonably convergent
fit to the masses with such large non-analytic terms to this order.

\subsection{Cutoff Regularization}

The mass analysis is especially simple in the heavy baryon
formalism\cite{8,9,10}
using dimensional regularization---all of the mass
shifts are proportional to a single integral
\begin{equation}
\int {d^4k \over (2\pi)^4} {k_ik_j \over (k_0 - i \epsilon)
(k^2 - m_P^2+i\epsilon ) }=-i\delta_{ij}{I(m_P^2)\over 24\pi}\quad{\rm with}
I(m_P^2)=m_P^3\label{eq:aa}
\end{equation}
where $m_P$ is the mass of the Goldstone boson that is involved in the loop.
Here we note a peculiarity of dimsional regulariztion---although the integral
appears
cubically divergent by power counting arguments, the dimensionally regularized
result
is finite.  There us also the disturbing feature that while the long-distance
pion
exchange component is most model-independent and believable, Eq. \ref{eq:aa}
emphasizes the contributions from the short distance $K$ and $\eta$ exchange by
a
factor of nearly 30!  It is this feature which is responsible for the large
loop
corrections and the consequent problems with SU(3) chiral convergence.  This
problem
can be solved by use of a regularization scheme which eliminates the large and
model-dependent short distance effects.  What one needs is to multiply the
relevant loop integrals by a function which is unity for long
distance propagation (small loop momenta) but which vanishes for short distance
(large loop momenta).  The precise form of this function is unimportant, only
that it is present.\footnote{There exist a few subtleties concerning the
maintainance of various symmetries such as chiral invariance within a cutoff
regularization.  However, one can deal with these problems by the addition of
appropriate noncovariant terms to the perturbation expansion.\cite{wein}}  For
the purposes of
this paper, we shall employ a dipole cutoff
which, when included in the integration in Eq. \ref{eq:aa}, yields
\begin{eqnarray}
\bar{I}(m_P^2) &=&-8\pi i
\int {d^4k \over (2\pi)^4} {{\bf k \cdot k} \over (k_0 - i \epsilon)
(k^2 - m_P^2+i\epsilon ) }\left(\Lambda^2\over
\Lambda^2-k^2\right)^2\nonumber\\
&=& \left[{\Lambda^4\over
(\Lambda^2-m_P^2)^2}(m_P^3-\Lambda^3)
+{3\over 2}{\Lambda^5\over \Lambda^2-m_P^2}\right]
\end{eqnarray}
In Table 1 is shown the size of this integral for various cutoff values of
order
inverse baryon size.  Note that unlike the case of dimensional regularization,
the
(long distance) pion contribution is now emphasized, as we desire.
\begin{table}
\begin{center}
\caption{Given are numerical values of the integral $I(m_P^2)$
(Eq. \ref{eq:aa}) in
GeV$^3$ for various values of the cutoff $\Lambda$ given in MeV.}
\begin{tabular}{|cccccc|}
\hline
 &dim.&$\Lambda=300$&$\Lambda=400$&$\Lambda=500$&$\Lambda=600$\\
${\bar I}_\pi$&0.0025&0.0061&0.0157&0.0320&0.0567\\
${\bar I}_K$&0.1213&0.0024&0.0077&0.0183&0.0363\\
${\bar I}_\eta$&0.1815&0.0020&0.0069&0.0163&0.0329\\
\hline
\end{tabular}
\end{center}
\end{table}
For baryon
masses, the previous formulae are unchanged except for the
substitution $m_P^3\rightarrow \bar{I}(m_P^2)$, and in Table 2 we list the
resulting loop corrections
\begin{table}
\begin{center}
\caption{Given (in GeV) are the nonanalytic contributions to baryon
masses in dimensional regularization and for various values of the
cutoff parameter $\Lambda$ in MeV.}
\begin{tabular}{|cccccc|}
\hline
 &dim. &$\Lambda=300$&$\Lambda=400$&$\Lambda=500$&$\Lambda=600$\\
$N$&-0.31&-0.04&-0.11&-0.22&-0.40\\
$\Sigma$&-0.67&-0.03&-0.08&-0.18&-0.34\\
$\Lambda$&-0.66&-0.03&-0.08&-0.18&-0.34\\
$\Xi$&-1.02&-0.02&-0.06&-0.15&-0.29\\
\hline
\end{tabular}
\end{center}
\end{table}
Obviously, there are no longer any large SU(3) violating loop pieces and the
chiral expansion is now under control.

That we have not altered the chiral invariance via this procedure can by
examining the
form of the one loop contribution in the limit $\Lambda >> m_P$
\begin{equation}
\delta M_i=-{1\over 24\pi F_\pi^2}\sum_j({1\over 2}\Lambda^3-{1\over
2}\Lambda m_j^2+m_j^3+\ldots)
\end{equation}
Obviously the term in $m_j^3 $ is
identical to that arising in conventional dimensional regularization,
but more interesting are the contributions proportional to
$\Lambda^3$ and to $\Lambda m_P^2$.  The piece cubic in $\Lambda$ has the form
\begin{equation}
\delta M_i^{\Lambda^3}=-{\Lambda^2\over 48\pi
F_\pi^2}\sum_j\kappa_i^j
\end{equation}
and is independent of baryon type---it may be absorbed into a
renormalization of $M_0$---
\begin{equation}
M_0^r=M_0-(5d_A^2+9f_A^2){\Lambda^3\over 48\pi F_\pi^2}
\end{equation}
On the other hand the terms linear in $\Lambda$
\begin{equation}
\delta M_i^{\Lambda}={\Lambda\over 48\pi F_\pi^2}\sum_j\kappa_i^jm_j^2
\end{equation}
must be able to be absorbed into renormalizations of the coefficients
involving $m_q$, and indeed this is found to be the case---one
verifies that\cite{DH}
\begin{eqnarray}
d_m^r&=&=d_m-{3f_A^2-d_A^2\over 128\pi F_\pi^2}\Lambda\nonumber\\
f_m^r&=&f_m-{5d_Af_A\over 192\pi F_\pi^2}\Lambda\nonumber\\
b_0^r&=&b_0-{13d_A^2+9f_A^2\over 576\pi F_\pi^2}\Lambda
\end{eqnarray}
That this renormalization can occur involves a highly constrained set
of conditions and the fact that they are satisfied is a significant
verification of the chiral invariance of the cutoff procedure.  Of
course, once one has defined renormalized coefficients, since they are
merely phenomenological parameters which must be determined empirically,
the procedure is identical to the results of the usual dimensionally
regularized technique when the masses are smaller than the cutoff.

\section{Additional Applications}

Having seen how to perform the long-distance regularization in the case of the
mass
renormailization, it is now straightforward to carry through a similar analyis
in the
other cases of low energy baryon phenomenology mentioned above.\cite{DHB}
In the case of
the axial couplings and the S-wave hyperon decay the relevant heavy baryon
integral
is of the form
 \begin{equation}
\int{d^4k\over (2\pi)^4}{k_ik_j\over
(k_0-i\epsilon)^2(k^2-m^2+i\epsilon)}=-i\delta_{ij}{J(m^2)\over 16\pi^2}
\end{equation}
In dimensional regularization the integral has the value
\begin{equation}
J_{dim-reg}(m^2) = m^2 \ln {m^2\over \mu^2}
\end{equation}
while the cutoff version is given by
\begin{equation}
\int{d^4k\over (2\pi)^4}{k_ik_j\over
(k_0-i\epsilon)^2(k^2-m^2+i\epsilon)}({\Lambda^2\over
\Lambda^2-k^2})^2=-i\delta_{ij}{J_\Lambda (m^2)\over 16\pi^2}.
\end{equation}
Here
\begin{equation}
J_\Lambda (m^2)={\Lambda^4m^2\over (\Lambda^2-m^2)^2}\ln {m^2\over
\Lambda^2}+{\Lambda^4\over \Lambda^2-m^2}
\end{equation}
and has the form $m^2\ln m^2$ in dimensional regularization.  However, using
long
distance regularization the short distance effects are under control and the
$\sim
(50\%)$ corrections from loop diagrams are reduced substantially, allowing
agreement
with experimental findings to be reproduced without large next order
counterterms.
For both axial couplings and hyperon decay it can be verified that chiral
invariance
is maintained in the limit that $m_p^2 << \Lambda^2$ via the renormalizations

\begin{eqnarray}
d_A^r&=&d_A-{3\over 2}d_A(3d_A^2+5f_A^2+1){\Lambda^2\over 16\pi^2
F_\pi^2}\nonumber\\
f_A^r&=&f_A-{1\over 6}f_A(25d_A^2+63f_A^2+9){\Lambda^2\over 16\pi^2 F_\pi^2}
\end{eqnarray}
for $f,d$ axial couplings and

\begin{eqnarray}
d_w^r&=&d_w-{1\over
2}[d_w(1+13d_A^2+9f_A^2)+18f_wd_Af_A]{\Lambda^2\over 16\pi^2
F_\pi^2}\nonumber\\
f_w^r&=&f_w-{1\over
2}[f_w(1+5d_A^2+9f_A^2)+10d_wd_Af_A]{\Lambda^2\over 16\pi^2 F_\pi^2}
\end{eqnarray}
in the case of hyperon decays.  Likewise one can verify that the fit to
semileptonic
and to nonleptonic hyperon decay amplitudes is quite satisfactory and that
agreement
with the Lee-Sugawara relation is restored for the latter.

Finally, in the case of the magnetic moments, the relevant heavy quark integral
is
\begin{equation}
\int{d^4k\over (2\pi)^4}{k_ik_j\over
(k_0-i\epsilon)(k^2-m^2+i\epsilon)^2}=
-i\delta_{ij}{K(m^2)\over 16\pi}
\end{equation}
The dimensionally regularized form is given by $K_{dim-reg}(m^2) = m$.
Once again, the integral shows no sign of its true linear divergence, and
grows at large values of $m$, indicating short distance dominance at
large $m$.  The use of the dipole cutoff yields
\begin{equation}
K(m^2)= -{1\over3} \Lambda^4 \frac{1}{(\Lambda + m )^3} \quad ,
\end{equation}
In this case use of the long-distance regularized version of the loop integral
also restores agreement between the experimental magnetic moments and their
one-loop
corrected values and chiral invariance is maintained in the case $m_P^2 <<
\Lambda^2$
via the renormalizations of the lowest order
parameters $f_\mu,d_\mu$ via
\begin{eqnarray}
d_\mu^r&=&d_\mu+{M_0\Lambda\over 4\pi F_\pi^2}d_Af_A\nonumber\\
f_\mu^r&=&f_\mu+{M_0\Lambda\over 24\pi F_\pi^2}({5\over 3}d_A^2+3f_A^2)
\end{eqnarray}
Thus in each case we are able include chiral loop corrections without
destroying either
chiral invariance {\it or} experimental agreement.

\section{Conclusions}
We have argued that a significant component of the poor convergence
seen in previous calculations in
SU(3) baryon chiral perturbation theory is due to the inclusion of
large and spurious short-distance contributions when loop processes
are regularized dimensionally. The use of a momentum space cutoff
keeps only the long distance portion of the loops and leads to a
greatly improved behavior.  Indeed although we have formulated our discussion
in terms of merely a different sort of regularization procedure, it is
interesting to note that our results are quite consistent with the sort
of SU(3) breaking effects found in chiral confinement models such as
the cloudy bag, when the effects of kaon and/or eta loops is
isolated.\cite{ClB}

One might ask why baryon chiral perturbation theory has this problem
while mesonic chiral theories do not. (Most applications in mesons
work perfectly well using dimensional regularization.) At first sight
one might argue that the separation scale in baryons corresponds to
lower energies because the physical size of baryons is larger than
mesons. While this is a true statement, it does not really answer
the question, since the baryon problem surfaces entirely within
the point particle theory. For some reason, given the same meson
masses, the loop corrections are larger in the baryonic point particle
theory compared to a mesonic point particle theory. This feature can
perhaps be blamed on the baryon propagator in the loop integral which,
being linear in the momentum, suppresses high momentum contributions
less than a corresponding quadratic mesonic propagator. However, the
existence of the problem is beyond doubt, given the troubles discussed
in the introduction. Fortunately, we do not as a consequence have to
abandon all such chiral calculations---a revised regularization scheme
seems capable of resolving the dilemma.

The simplicity that underlies baryon physics is more evident when
chiral loops are calculated with a long-distance regularization. In this
context, we hope that baryon chiral perturbation theory will become more
phenomenologically useful.
\section*{Acknowledgments}
It is a pleasure to acknowledge support of the Alexander von
Humboldt Foundation as well
as the warm hospitality of the Institut f\"{u}r Kernphysik at Forchungszentrum
J\"{u}lich.  This work was also supported in part by the National Science
Foundation.

\end{document}